# Second-Order Jahn–Teller Distortions and Dynamic Lattice Polarizability as the Origin of Broadband Emission in Bi$^{3+}$-Doped Cs$_2$SnCl$_6$

Shruti Prasad[#], Pabitra Kumar Nayak[#], Dibyajyoti Ghosh[*]

## Abstract

Lead-free vacancy-ordered perovskites (VOHPs) such as Cs$_2$SnCl$_6$ have emerged as promising materials for optoelectronic applications but typically suffer from wide band gaps and low photoluminescence quantum yields (PLQY). In this work, the electronic origins of broadband blue emission in Bi$^{3+}$-doped Cs$_2$SnCl$_6$ are elucidated by combining density functional theory (DFT) calculations and *ab initio* molecular dynamics simulations. The study systematically explores the spatial configurations of Bi$^{3+}$ dopants and Cl$^-$ vacancies, assessing their impact on structural and electronic properties. Results demonstrate that the formation of square pyramidal BiCl$_5^{2-}$ serves as efficient exciton traps, fundamentally enabling strong luminescence in an otherwise non-emitting host. The Bi$^{3+}$ 6s$^2$ lone pair in an asymmetric coordination environment drives second-order Jahn–Teller distortions, creating a highly polarizable local lattice. This softness, in contrast to the rigid [SnCl$_6$]$^{2-}$ framework, produces heterogeneous distributions of bond lengths, bond angles, and band edges around dopant sites. Defect-induced states strongly localize charge carriers, while coupling to vibrational modes at 300 K dynamically broadens the band edges, accounting for the large Stokes shift and broadband PL. These results establish lone pair–driven Jahn–Teller distortions and lattice heterogeneity as key design principles for efficient luminescence in lead-free perovskites.

## Introduction

Lead halide perovskites have attracted the attention of the material science community for optoelectronic applications for a long time due to their high absorption coefficient, high photoluminescence quantum yield (PLQY, ≈90%), narrow full width at half maximum (FWHM = 12–25 nm) [1], tuneable band gap [2] and shallow defect states [3]. These properties have enabled their use in a wide range of functional applications, such as solar cells [4], light-emitting diodes (LEDs) [5], photocatalysis [6], and photodetection [7]. However, the high water solubility of lead-based perovskites and the adverse effects of long-term lead exposure on living organisms make the search of non-toxic alternatives imperative [8,9].

Vacancy-ordered perovskites (VOPs) have emerged as one of the best-suited candidates due to the increased stability of +4 oxidation state of alternative B-site cations like Sn, Ti, Zr, Hf, Te, Pd, Pt compared to their +2 oxidation states found in conventional ABX$_3$ perovskites [10,11]. The VOP A$_2$BX$_6$ structure can be viewed as a derivative of the ABX$_3$ structure, in the sense that the A$_2$BX$_6$ structure is obtained by repeating the ABX$_3$ structure in each crystallographic direction and removing alternative B-site cations. This leads to the formation of isolated [BX$_6$]$^{2-}$ octahedra, eliminating B–X–B connectivity, resulting in the valence band maximum (VBM) being derived solely from the halogen orbitals compared to the ABX$_3$ type perovskites whose VBM has contributions from both, the B-site cation, and the halogen [11]. The band gap of VOPs, thus becomes more susceptible to the effect of halogen substitutions and halide-related defects. The isolated nature of the structure allows for a broader range of compositional tuning, as B-site cations and X-site anions with a wider range of ionic radii can be accommodated [12]. In addition, the shorter B–X bond length further contributes to the enhanced structural stability of these materials.

Vacancy-ordered perovskites, though more stable and environmentally benign, typically exhibit wider band gaps and low photoluminescence quantum yields (PLQY) due to their isolated octahedral structure. To engineer the VOP material class for different optoelectronic applications, several compositional strategies have been explored. Cation-based alloying (e.g., Cs$_2$Sn$_{1-x}$Te$_x$Cl$_6$ [13,14]) yields very high PLQY and enhanced moisture and thermal stability, Cs$_2$Ti$_{1-x}$Sn$_x$X$_6$ where X = I, Br [15] enables tuneable optical properties and improved structural stability, and Cs$_2$Zr$_{1-x}$Te$_x$Cl$_6$ [16] demonstrates enhanced PLQY, etc). Halogen-based alloying (e.g., Cs$_2$SnI$_{6-x}$Br$_x$ [17] and Cs$_2$TiI$_{6-x}$Br$_x$ [18]) effectively tunes the band gap while enhancing thermal stability. Tri-valent B-site metal doping (Bi$^{3+}$, Sb$^{3+}$) [19-25] introduces strong luminescence, whereas co-doping (Bi$^{3+}$ and Te$^{4+}$) [21] enables tuneable colour mixing. The introduction of defects such as Sn$^{2+}$ substitution and halogen vacancies [25] further enhances photoluminescence efficiency.

Pristine Cs$_2$SnCl$_6$ exhibits poor photoluminescence property, however, due to its zero-dimensional structure, it is an excellent host material as it can localize the excitons at the luminescent centres, thereby mitigating non-radiative recombination and enabling high-efficiency emissions.[25] By doping Cs$_2$SnCl$_6$ with $Bi^{3+}$ ions, Tan et al. [22] were able to achieve a direct band gap of ≈3.0 eV compared to $Cs_2SnCl_6$ (3.9 eV) and an impressive PLQY = 78.9% (emission peak: 455 nm with a wide FWHM of 66 nm and a large Stokes shift of 90 nm).

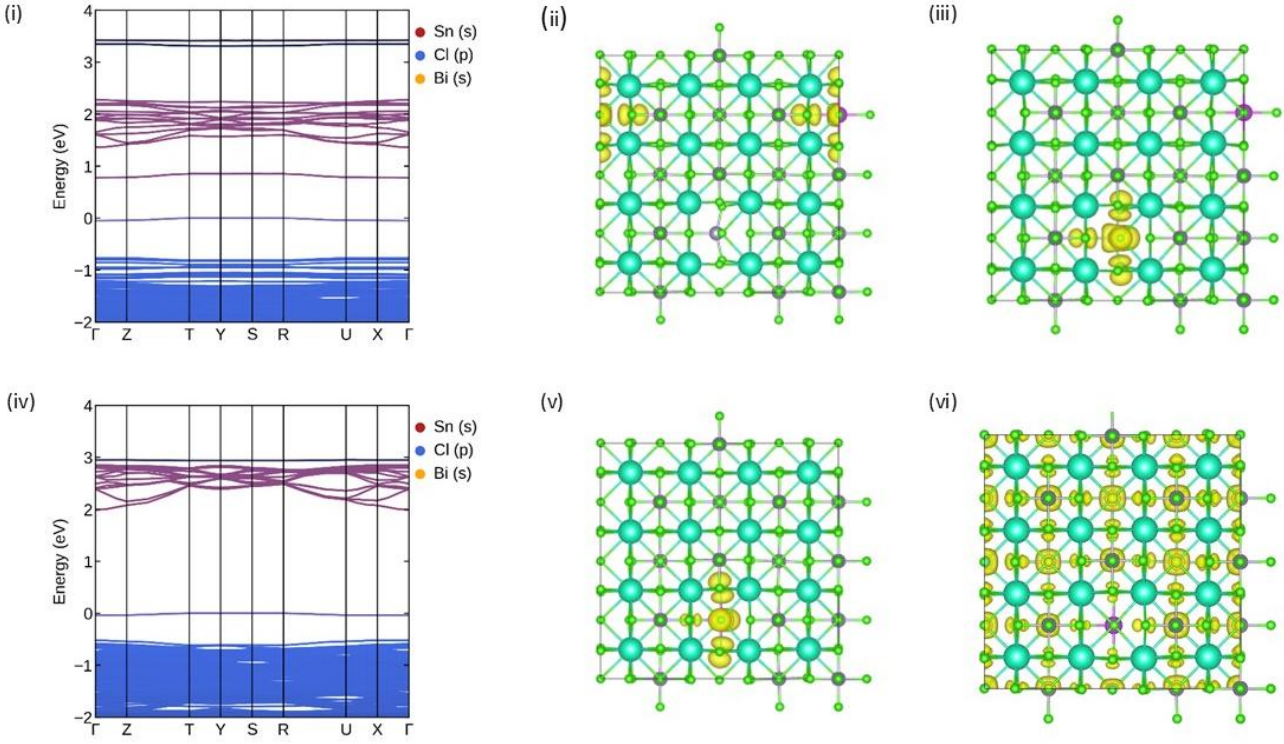

*Figure 1: (i)-(iii) DFT calculated band structure and band decomposed charge densities of VBM'(ii) and CBM' (iii) of isolated defects configuration. (iv)-(vi) DFT calculated band structure and band decomposed charge densities of VBM'(ii) and CBM (iii) of clustered defects configuration.*

Doping in $Cs_2SnCl_6$ leads to a disruption in charge neutrality, which the system compensates for either by trapping electrons and holes in localized regions or by introducing additional electronic states within the bandgap [26, 27]. These localized states influence the material's luminescent behaviour. Tan et al. attributed the reduction in band gap of the doped system to the formation of a defect state above the VBM of the bulk, which they called VBM'. They hypothesize the reason behind the high PLQY to be the trapping of excitons by the $[BiCl_5]^{2-}$ pyramid which are strongly confined in the 0D $Cs_2SnCl_6$ perovskites. They recorded the maximum PLQY for the sample having Bi concentration of 2.75%. The reason for degradation of PLQY at higher concentrations was proposed to be the clustering of $[BiCl_5]^{2-}$ pyramids reducing the localization.

Since the study by Tan et al, several investigations have explored the electronic origins of the exceptionally high PLQY in $Bi^{3+}$-doped $Cs_2SnCl_6$. However, the fundamental mechanism responsible for the broadband emission and the reduction in PLQY at higher dopant concentrations remains unresolved. In this study, we systematically examine multiple defect configurations to identify and confirm the nature of the luminescent centres. We further model the increase in $Bi^{3+}$ concentration by varying the proximity between dopant sites to elucidate the origin of PLQY degradation at higher doping levels. Finally, we perform *ab initio* molecular dynamics simulations to probe the dynamic structural and electronic behaviour, thereby uncovering the mechanism underlying the enhanced PLQY and the experimentally observed broadband blue emission.

## Results and Discussion

DFT calculations using the HSE–SOC functional were performed for pristine $Cs_2SnCl_6$, Bi-doped systems with $Cl^-$ vacancies arranged in clustered and isolated configurations, and supercells with two $[BiCl_5]^{2-}$ octahedra at varying separations. The computationally optimised structures retained their experimentally observed geometry i.e., F*m*-3m. An optimized cell parameter of 10.425 Å was obtained for pristine $Cs_2SnCl_6$. The doped systems show an increase in lattice parameter up to 0.92% with the maximum lattice parameter being 10.521 Å. The substitution caused negligible impact on the static internal structure of the $SnCl_6$ octahedra.

## Static Electronic Properties: Effect of location of $Cl^-$ vacancy

To study the effect of location of $Cl^-$ vacancy we performed DFT calculations on two Bi-doped $Cs_2SnCl_6$ supercells: one in which the $Bi^{3+}$ dopant and $Cl^-$ vacancy are located within the same $BX_6$ octahedron (forming a defect cluster), and another where they are separated across different $BX_6$ octahedra, with a distance of 29.797 Å between the two B-sites. The supercell with isolated defects (in different octahedra) exhibited more pronounced local structural distortions due to isolation of effective charges, resulting in a 0.58% increase in cell volume, compared to only a 0.31% increase in the clustered configuration. Interestingly, the total energy of the supercell with isolated defects was slightly lower (–4.125 eV/atom) than that of the clustered defect system (–4.119 eV/atom).

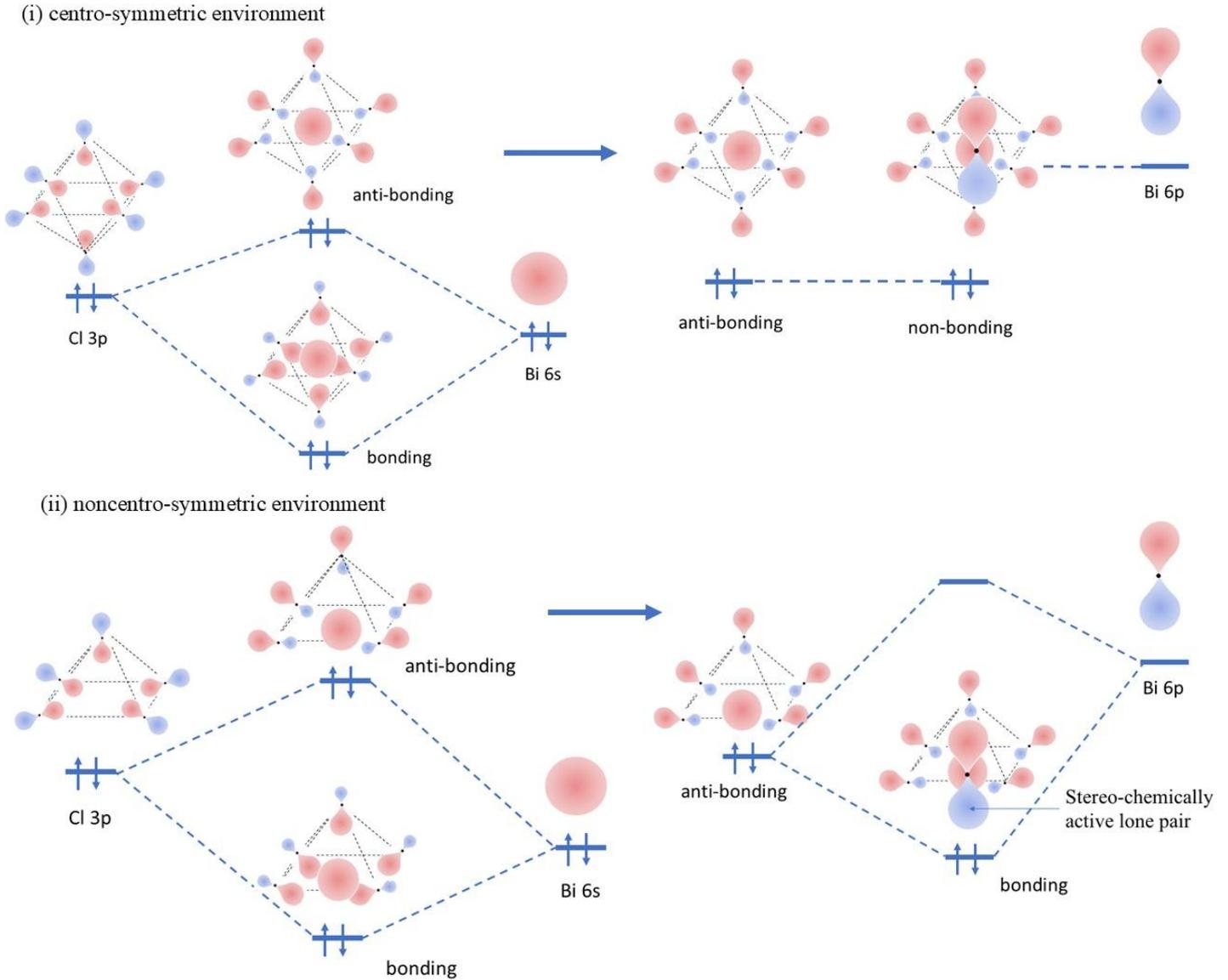

*Figure 2: Bonding in an octahedron vs a pyramid highlighting the difference in the p orbital coupling. In case of a cento-symmetric environment (i), destructive interference results in no bonding between Bi 6p and Bi 6s - Cl 3p antibonding state, where as, in case of asymmetric environment (ii), coupling between Bi 6p and Bi 6s - Cl 3p antibonding state results in a lone pair at the B-site.*

Using the HSE-SOC functional, direct band gaps of 3.023 eV and 1.750 eV were obtained for the supercells with clustered and isolated defects, respectively, in contrast to the pristine $Cs_2SnCl_6$ band gap of 3.848 eV. As shown in Fig. 1(i) and (iv), both the doped systems exhibit a flat band above the valence band maximum which we will call VBM′. The system with isolated defects also has a flat band below the conduction band minimum which we will call CBM′. Due to the existence of the VBM′ flat band, both the doped systems have very heavy holes, while the system having isolated defects has very heavy electrons as well due to the CBM′ flat band.

The band-edge charge densities and the projected density of states show that both the bulk VBM and the bulk CBM are delocalized over the Sn-Cl framework, without any identifiable contributions from the Cs atoms. The VBM has contributions from non-bonding Cl 3p orbitals, whereas the contributions to the CBM comes from σ-antibonding overlap between the 3p orbitals of Cl and 5s orbitals of Sn. The VBM′ band is primarily derived from the σ-antibonding overlap of 6s orbital of the dopant $Bi^{3+}$ cation and 3p orbitals of the $Cl^-$ anions in the corresponding octahedron while CBM′ band is derived from the σ-antibonding overlap of Sn 5s and Cl 3p orbitals of the octahedron containing the $Cl^-$ vacancy. There are several shallow defect states localized to the Bi-containing octahedra present near the VBM in case of the isolated defects configuration, while the defect cluster configuration has only one shallow defect state between VBM and VBM′.

In the case of isolated defects, VBM' primarily derived from Bi 6s and Cl 3p is lower in energy than bulk CBM,

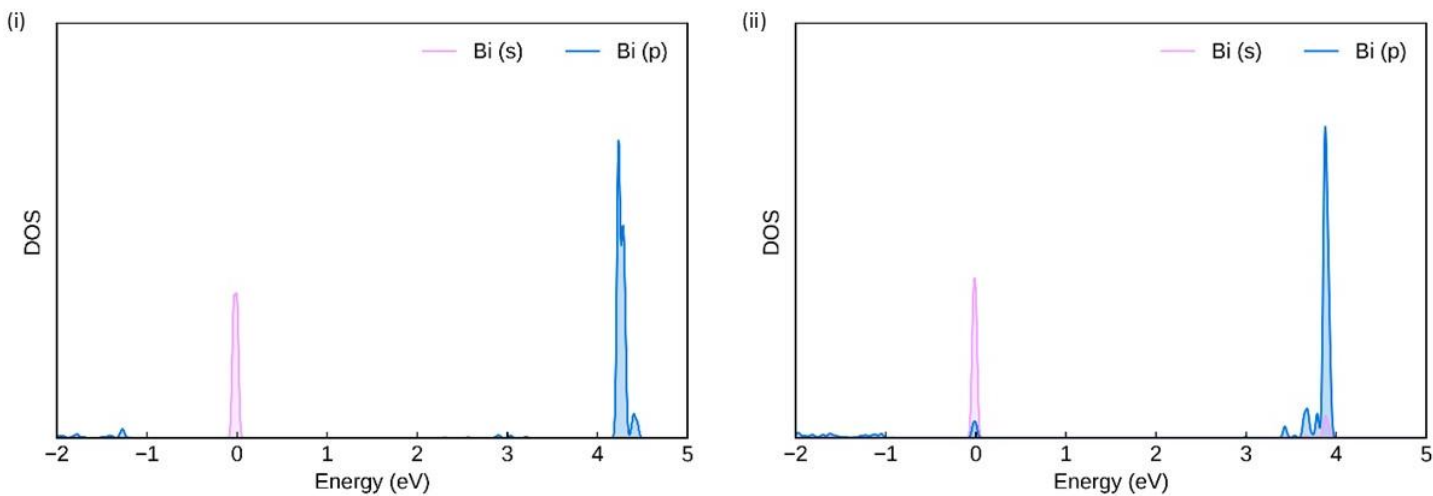

*Figure 3: (i) Density of State of Bi cation is isolated defects configuration, (ii) Density of State of Bi cation is clustered defects configuration highlighting the difference in p orbital coupling.*

which originates from Sn 5s and Cl 3p orbitals, because the Bi 6s states have much lower energy than Sn 5s states. Additionally, the defect-induced conduction band minimum (CBM′) is also lower than the bulk CBM, as it involves contributions from only five Cl$^-$ anions instead of six. For clustered defects, the combined influence of the deeper Bi 6s orbitals and reduced contributions from Cl$^-$ anions results in VBM' being even lower than that of the isolated defect case.

We also calculated a qualitative estimate of the transition dipole moment associated with the radiative recombination responsible for photoluminescent emission, i.e., the transition from the defect-derived VBM′ to the CBM (in the defect-clustered system) or to the CBM′ (in case of with isolated defects). The defect cluster shows a stronger optical transition with a transition dipole moment of 0.137 Debye, nearly an order of magnitude larger than the isolated defect case (0.0159 Debye). This enhancement in dipole strength translates into a higher oscillator strength and a much shorter radiative lifetime (1.9 ns vs 29.6 ns), indicating significantly increased transition probability. Therefore, the transition from VBM′ to CBM in the clustered defect configuration resulting in a direct band gap of 3.023 eV is much more likely to contribute to the experimentally observed photoluminescence even though the formation energy of isolated defects is similar, even slightly lower than that for the defect cluster. This matches well with the photoluminescence peak observed at 454 nm by Tan et. al.

Another significant observation evident from the band-edge charge densities of bands having contribution from octahedra containing the Cl$^-$ vacancy is the difference in the position and charge distribution of halide ions in case of having Bi or Sn at the B-site. The isosurface of VBM′ of the clustered defect configuration shows depletion of charge on the Cl$^-$ ion opposite the vacancy with the other Cl$^-$ ions slightly titled away from the vacancy. This is unusual since the remaining Cl$^-$ ions must have come closer to the vacancy to relax the electrostatic forces, which happens for the $[SnCl_5]^{1-}$ octahedron, along with similar charge density on each Cl$^-$ ion. This difference is caused by the presence of stereo-chemically active lone pair of electrons in Bi 6s orbital in the +3 oxidation state while Sn 5s orbital is empty in the +4 oxidation state. Coupling between the Bi 6p orbitals and antibonding state derived from Bi 6s and Cl 3p further results in bonding and non-bonding states, the bonding state becomes VBM'. As shown in Fig. 2, in case of asymmetric coordination environment p orbital coupling leads to the B-site having a lone pair, while in case of centro-symmetric environment this orbital interaction is suppressed due to destructive interference, as can be observed for $[BiCl_6]^{3-}$ octahedron, which also leads to narrowing of the band gap. The presence of a lone pair at the B-site becomes more important for dynamic fluctuations in the material at 300K.

## Static Electronic Properties: Effect of proximity of dopants

Next, to study the effect of increasing the dopant concentration and varying the distance between dopants, we performed DFT calculation for supercells having two luminescent centers $[BiCl_5]^{2-}$ at a distance of 10.515 Å, 12.779 Å and 15.517 Å respectively. The average total energy across the three relative dopant locations (-4.125 eV/atom) is equivalent to the pristine VOP, indicating the isolated $BX_6$ octahedral structural is able to accommodate the local lattice distortions due to substituting a larger $Bi^{3+}$ cation in place of a smaller $Sn^{4+}$ cation. A small increase in cell volume in the range of 0.55 – 0.57% is observed.

For systems with $Bi^{3+}$ dopants existing as a $[BiCl_5]^{2-}$ unit, a slight band gap widening is observed with increasing dopant concentration. At 6.25% Bi, the gap is 3.023 eV, while at 12.5% Bi it increases modestly to 3.027–3.089 eV, with larger Bi–Bi separations yielding slightly higher values due to reduced coupling. The optical gap remains essentially invariant, confirming that the fundamental transition energy is insensitive to dopant proximity and that the system retains a quasi-direct character (variation ≤2% across separations). By contrast, the band-edge curvature shows pronounced sensitivity: the hole effective mass increases dramatically with decreasing Bi–Bi distance, from ~25 $m_0$ to ~146 $m_0$, signalling a flattened VBM and strong hole localization near closely spaced Bi pairs. Electron masses, in contrast, remain moderate and relatively stable. The strong defect-trapping of holes at high Bi-Bi proximity leads to non-radiative recombination explaining the reduction in PLQY above Bi concentration of 2.75% observed by Tan et. al.

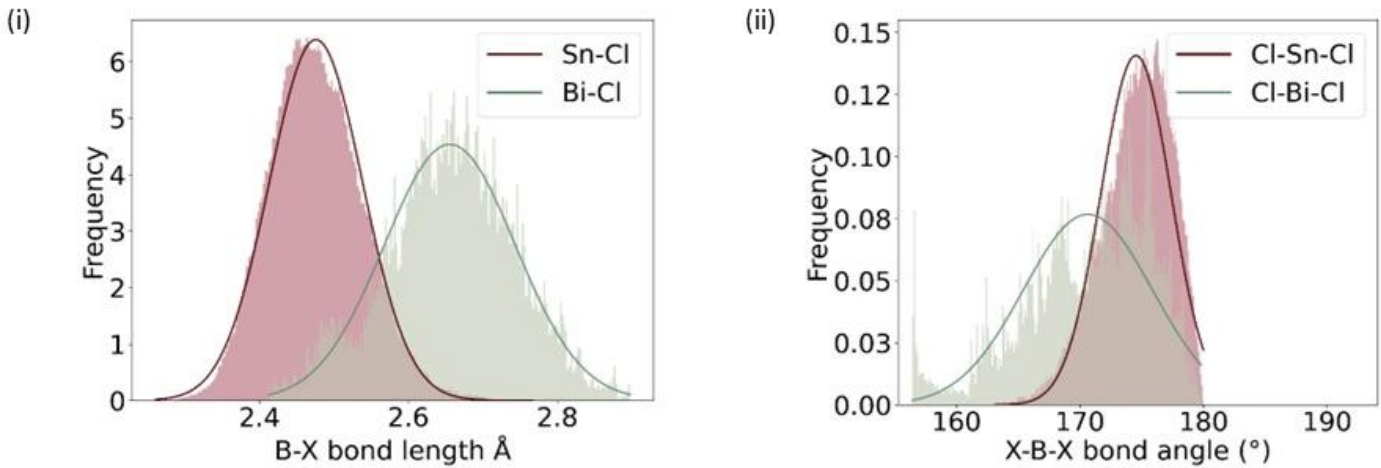

Figure 5: Distribution of (i) B-X bond length and (ii) X-B-X bond angle over the 3ps trajectory at 300K

Table 1: DFT obtained optical band gap and effective masses at different dopant concentrations and positions.

| Bi Conc-entration | Distance between $Bi^{3+}$ cations (Å) | Band gap (eV) | | $m_h$ ($m_0$) | $m_e$ ($m_0$) |
|---|---|---|---|---|---|
| | | Direct (eV) | Indirect (eV) | | |
| 0 | - | 3.848 | | -5.910 | 0.621 |
| 6.25% | - | 3.023 | 2.993 | | 0.982 |
| 12.5% | 10.515 | 3.027 | 3.023 | -24.702 | 0.832 |
| 12.5% | 12.779 | 3.067 | 3.046 | -15.804 | 0.968 |
| 12.5% | 15.517 | 3.089 | 3.112 | -146.117 | 0.745 |

The band-edge charge densities and the partial density of states exhibit the same lattice-site and elemental orbital contributions to VBM, VBM′ and CBM as the contributions to bands in case of single Bi doping. Even when the two dopants are placed at adjacent lattice positions, the band-edge charge density of VBM′ shows no overlap between the charge clouds at the two $[BiCl_5]^{2-}$ pyramid. This demonstrates that the octahedral units are electronically decoupled, highlighting the strong quantum confinement effect.

## Dynamic Structural Properties

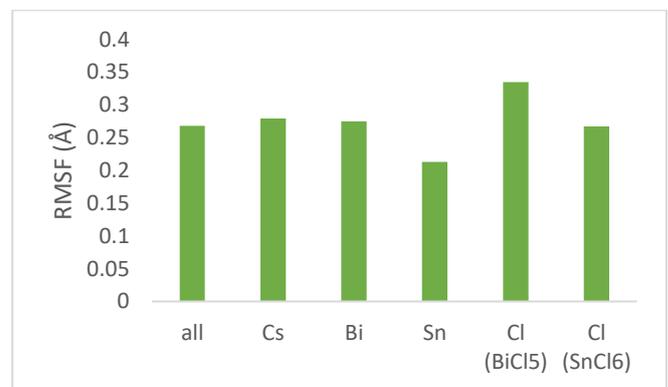

Figure 4: RMSF of different elements for 3ps trajectory at 300K

Finally, to examine the origin of the broadband emission (FWHM of 66 nm), we performed *ab initio* molecular dynamics (AIMD) simulations at 300 K to probe the dynamic behaviour of Bi-doped $Cs_2SnCl_6$. The root mean square fluctuation (RMSF) was calculated for each element to quantify the extent of thermally induced fluctuations in the lattice. With an overall RMSF of 0.268 Å, the material emerges as highly stable under thermal conditions.

RMSF analysis, Fig. 4, reveals pronounced dynamical heterogeneity. The defect-free $[SnCl_6]^{2-}$ framework remains comparatively rigid (Sn: 0.213 Å; $Cl^-$: 0.267 Å), whereas the $[BiCl_5]^{2-}$ unit exhibits significantly larger halide fluctuations ($Cl^-$: 0.335 Å, ~25% higher). This enhanced local softness reflects weaker bonding in the $[BiCl_5]^{2-}$ pyramid compared to the $[SnCl_6]^{2-}$ octahedron, arising from the longer Bi–Cl bond length due to the larger ionic radius of $Bi^{3+}$ and the presence of a $Cl^-$ vacancy and a lone pair at the B-site, which make the potential anharmonic.

To further study the dynamic effect of B-site doping with a heterovalent cation, we measure the thermal fluctuations of the structural properties that affect the optoelectronic performance of the material. Fig. 3 shows histogram plots of the B-X bond length, the X-X distance, the X-B-X angle, and the X-B-B-X dihedral angle for B = Sn and Bi, while Table 2 has their mean and standard deviations respectively.

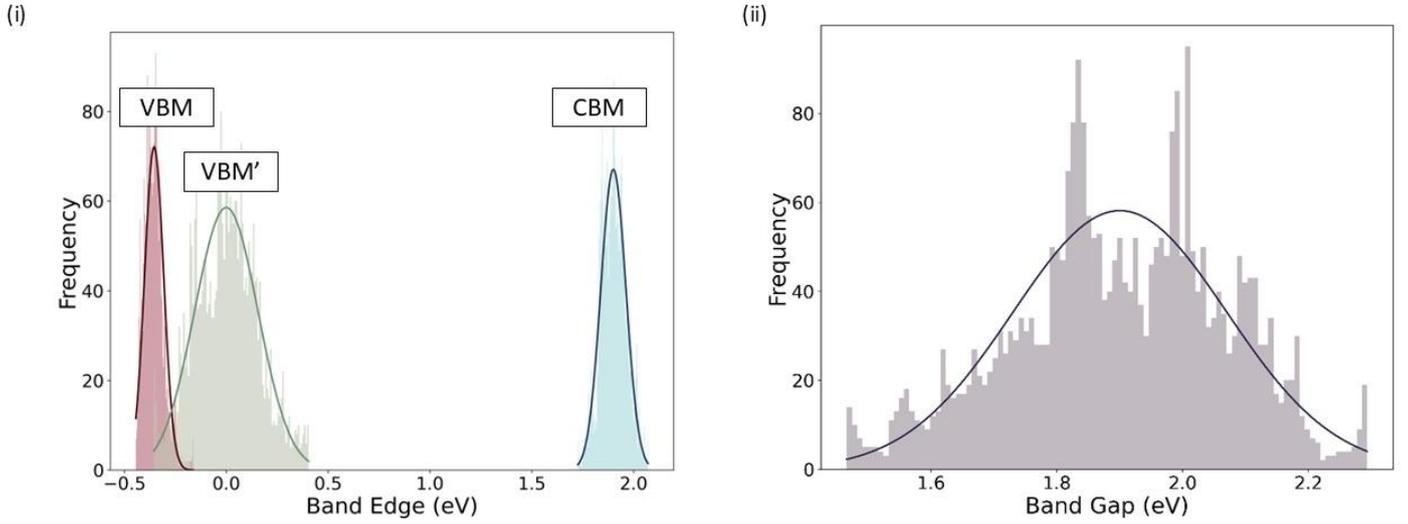

*Figure 6: DFT calculated (PBE) distribution of (i) VBM, VBM' and CBM, and (ii) Band Gap over the 3ps trajectory at 300K*

*Table 2: Mean and standard deviation of structural properties that influence the optoelectronic properties.*

| Structural Properties | μ | σ |
|---|---|---|
| Bi-Cl bond length | 2.655 Å | 0.088 Å |
| Sn-Cl bond length | 2.476 Å | 0.062 Å |
| Cl-Cl distance in $BiCl_5$ | 3.781 Å | 0.162 Å |
| Cl-Cl distance in $SnCl_6$ | 3.499 Å | 0.118 Å |
| Cl-Bi-Cl bond angle | 170.650° | 5.200° |
| Cl-Sn-Cl bond angle | 174.541° | 2.835° |
| Cl-Bi-Sn-Cl dihedral angle | -1.075° | 5.317° |
| Cl-Sn-Sn-Cl dihedral angle | 0.000° | 4.618° |

Because the relevant band edges in Bi-doped $Cs_2SnCl_6$ derive from σ-antibonding overlaps: VBM' from Bi 6s-Cl 3p and CBM from Sn 5s-Cl 3p, the B–X bond length is a primary structural lever for the electronic structure. Bi–Cl bonds are ~0.18 Å longer (~7.2%) and have 42% higher σ than Sn-Cl bonds, consistent with the larger ionic radius of $Bi^{3+}$ and the softer, vacancy-adjacent $[BiCl_5]^{2-}$ anharmonic environment. The longer Bi-Cl bond lead to lesser overlap and weaker bonding compared to Sn-Cl bonds also resulting in higher fluctuations. The average X-B-X bond angles calculated for the two B-site cations highlight the effect of the presence of a lone pair at Bi. The Cl-Bi-Cl angle is much lower ~4° compared to the Cl-Sn-Cl angle because the lone pair at Bi repels the $Cl^-$ anions resulting in second-order Jahn–Teller effects. Cl-Bi-Cl bond angle also has an 83% higher standard deviation compared to Cl-Sn-Cl bond angle due to the asymmetric coordination of $Bi^{3+}$ in the $[BiCl_5]^{2-}$ unit, where the lone pair drives off-centring and weakens the restoring force on the $Cl^-$ ions. The intra-octahedral Cl-Cl distances have same averages as geometrically calculated from the average B-X bond length. The X–B–B–X dihedral angle distributions for Bi and Sn containing units overlap significantly, with nearly identical averages and only a marginally larger spread for Cl–Bi–Sn–Cl. This indicates that substitution of Bi does not strongly perturb the cooperative tilting or rotation of adjacent octahedra.

## Dynamic Electronic Properties

*Table 3: Mean and standard deviation of band edges and Band Gap*

|  | μ (eV) | σ (eV) |
|---|---|---|
| VBM | -0.354 | 0.047 |
| VBM' | 0.00 | 0.156 |
| CBM | 1.900 | 0.061 |
| Band Gap | 1.900 | 0.170 |

To examine the influence of the finite-temperature structural dynamics on the optoelectronic properties of Bi doped $Cs_2SnCl_6$, the distribution of band edges and band gap is plotted in Fig.5. VBM' has a much wider distribution, Table 3, σ more than twice than that of CBM and four times more than that of VBM, clearly demonstrating how the introduction of $[BiCl_5]^{2-}$ defect transformed a non-emitting host to a material having a PLQY as high as 78.9% by introducing a fluctuating defect-state. The fluctuations in the band edges can be corelated to the structural dynamics as VBM' energy level directly depends on the extent of overlap between the Bi 5s and Cl 3p orbitals, which, in turn, depend on the Bi-Cl bond length. Similarly, VBM depends on the Sn-Cl bond length and CBM on the Cl-Cl distance. A highly fluctuating Bi-Cl bond length results in a widely distributed VBM'. As a result, the band gap has a wide distribution with σ = 0.170 eV, which matches well with the experimentally observed FWHM value of 66 nm for the PL.

Next, to examine whether the static localized or delocalized character of the band edges is preserved in the dynamic states, we plot the band decomposed charge density for a few snapshots that correspond to a peak or trough in VBM′. For all the snapshots, CBM, though less delocalized than the static state due to the thermal fluctuations, remains fairly delocalized over the Sn-Cl framework. For configurations corresponding to peaks in the VBM′ energy, VBM′ remains localized to the $[BiCl_5]^{2-}$ pyramid. However, for 40% of lower VBM′ energy configurations considered, VBM′ also has contributions from nearby octahedra. Even though VBM′ is slightly delocalized for some snapshots in the 3 ps trajectory, the extent of delocalization is very small, this is supported by an average Inverse Participation Ratio (IPR) value of 0.160 for VBM′ compared to a value of 0.031 for CBM.

To pinpoint the exact structural fluctuations that lead to such a wide distribution of band gap at 300K, we extracted the spectral density, Fig. 6, from the band edge energy distribution of the 3ps trajectory. The spectral density plot showed a strong peak around 50 – 75 cm$^{-1}$ and smaller peaks around 300 cm$^{-1}$. These peaks match well with experimental Raman spectroscopy results [32] obtained for $Cs_2SnCl_6$ with the peak around 50 – 75 cm$^{-1}$ corresponding to the vibrations of $Cs^+$ cation sublattice against the rigid $[BX_6]^{2-}$ octahedra, 232 cm$^{-1}$ one corresponding to B-X symmetric stretch and the one around 309 cm$^{-1}$ corresponding to B-X asymmetric stretch. The vibrational mode around 50 – 75 cm$^{-1}$ forms a phonon bath, energetically coupled to the local anharmonic environment surrounding the $[BiCl_5]^{2-}$ pyramid. Such coupling dynamically amplifies Bi–Cl bond-length and angle fluctuations during thermal motion, thereby promoting transient localization of carriers at the defect, as corroborated by the broad distribution of VBM′ energies and the experimentally observed broadband emission. The peaks around 232 cm$^{-1}$ and 309 cm$^{-1}$ prove the correlation between the Bi-Cl bond length and the band gap.

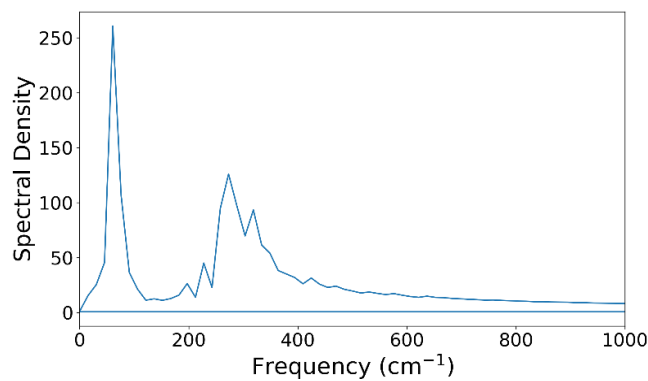

Figure 7: Spectral Density plot for the transition from VBM′ to CBM for the 3ps trajectory

## Conclusion

The DFT calculations for varying Bi doping configurations and concentrations confirm the formation of $[BiCl_5]^{2-}$ pyramids as luminescent centres. At low concentrations, these centres enhance photoluminescence, while at higher concentrations, modelled here as reduced Bi–Bi separation, the onset of extremely heavy hole effective masses quenches PL due to non-radiative recombination. Ab initio molecular dynamics at 300 K revealed enhanced lattice fluctuations upon defect introduction. The anharmonic potential within the defect octahedron, arising from the Cl$^-$ vacancy and the stereo-chemically active lone pair on Bi$^{3+}$, renders the $[BiCl_5]^{2-}$ pyramid highly polarizable compared to the rigid $[SnCl_6]^{2-}$ framework. This dynamical heterogeneity facilitates self-trapped exciton (STE) formation: upon photoexcitation, structural fluctuations at the defect site lower the trapping barrier and localize the exciton. The electronic decoupling of the rigid host octahedra supports high PLQY, while the fluctuating defect states stabilize trapped excitons, giving rise to broadband emission with a large Stokes shift.